Bose- Einstein condensation and the glassy state

Moshe Schwartz

School of Physics and Astronomy

Raymond and Beverly Sackler Faculty of Exact Sciences

Tel Aviv University, Ramat Aviv

Tel Aviv 69978, Israel

Abstract

The distinction between a classical liquid and a classical ordered solid is easy and depends on their different symmetries. The distinction between a classical glass and a classical liquid is more difficult, since the glass is also disordered. The difference is in the fact that a glass is frozen while the liquid is not. In this article an equilibrium measure is suggested that distinguishes between a glass and a liquid. The choice of this measure is based on the idea that in a system which is not frozen symmetry under permutation of particles is physically relevant, because particles can be permuted by actual physical motion. This is not the case in a frozen system. In quantum Bose systems there is a natural parameter that can distinguish between a frozen and a non-frozen state. This is the Bose condensed fraction. In this article it is shown how to generalize this concept, in a natural way, to classical systems in order to distinguish between the glass and the liquid. A formal similarity between the ground state wave function of a Bose system and the statistical weight function enables a definition of a Bose Einstein condensed fraction in classical systems. It is finite in the liquid and zero in the frozen state. The actual value of the condensed fraction in the liquid may serve also as a measure of the glassiness in the liquid. The smaller it is the larger the glassiness of the liquid.

The freezing of a liquid into an ordered solid is characterized by the appearance of Brag peaks in X ray scattering from the solid at momenta corresponding to the reciprocal lattice. This happens in addition to the latent heat and discontinuity of the density at the transition that are not specific to that transition. The strength of the Brag peaks is thus an equilibrium parameter that distinguishes between the liquid state and the state of a periodic solid. When dealing with a glass the situation is different, because the structure factor that is measured by X ray scattering is similar to that of a liquid. The usual measures that distinguish a liquid from a glass are various transport parameters such as the diffusion constant of a tagged particle in the glass or the viscosity of the glass versus the viscosity of the liquid [1,2]. Both measures are based on the fact that a glass is a frozen system while the liquid is not. In the following I'll suggest a measure that also distinguishes between the frozen and unfrozen states but is basically an equilibrium measure. The idea behind this suggestion is that in a liquid of **classical** identical particles the symmetry under permutations of particles is physically relevant. Namely, two configurations of the system that differ only in permutations of the particles can be connected by physical continuous trajectories in configuration space that do not have to pass through energy barriers that are too large (e.g. larger than the thermal energy) .In a glass, on the other hand , while symmetry under permutations still formally exists, it seems not to be relevant , because continuous paths in configuration space connecting two configurations that differ by a permutation which has at least one large cycle will have to cross large energy barriers. The natural parameter that distinguishes quantum systems in which the symmetry under permutations is physically relevant from systems where it is not relevant is the Bose condensed fraction. The ground state of a Bose system exhibits a finite condensed fraction while that of the corresponding Fermi system, in which the ground state is not invariant under permutations, the condensed fraction is zero. (I will not go here into the finite temperature description of the quantum mechanical system, since the analogy to be used here between the quantum mechanical and classical systems is between the quantum mechanical ground state and the classical distribution function.) Consider first this analogy. The following set of Langevin equation describes the stochastic over- damped motion of a classical system consisting of $N$ particles and enclosed in a cube of volume $V$ with periodic boundary conditions:

$$\frac{d}{dt}x_i^k = -\gamma \frac{\partial}{\partial x_i^k}W + \eta_i^k \ , \tag{1}$$

where $W$ is the potential energy of the system, $i$ denotes the particle , $k$ denotes the Cartesian component of a vector and the noise $\eta$ obeys

$$\langle \eta_i^k \rangle = 0 \qquad \text{and} \qquad \langle \eta_i^k(t)\eta_j^l(t')\rangle = 2D\delta_{ij}\delta_{lm}\delta(t-t') \ . \tag{2}$$

It is well known that the above Langevin equations can be transformed into a Fokker Planck equation for, $P$ , the distribution function in configuration space

$$\frac{\partial P}{\partial t} = \sum_i \nabla_i \cdot [D\nabla_i + \gamma \nabla_i W]P \ . \tag{3}$$

This has as steady state solution the Gibbs distribution $P_S \propto \exp[-W/kT]$, where $kT = D/\gamma$. A standard transformation $P = P_S^{1/2}\Psi$ [3], leads to an imaginary time Schrodinger equation

$$\frac{\partial \Psi}{\partial t} = -H\Psi. \tag{4}$$

The precise form of $H$ is not of interest in the following and will not be given here. It is important to note, however, that it is Hermitian, and non negative definite, where its only eigenstate with zero eigenvalue is the ground state $\Psi_G = P_S^{1/2}$. This defines for the classical system in thermal equilibrium a natural "ground state". Clearly the $\Psi_G$ is symmetric under permutations and therefore is the ground state of the Bosonic reduction of $H$. Therefore, a Bose condensed fraction $\xi$ can be defined for the classical liquid as the ground state condensed fraction of some quantum Bose system with a ground state given by $\Psi_G = P_S^{1/2}$. Clearly the ground state condensed fraction is a functional of the ground state. This functional is known for many years to be given by

$$\xi = \frac{1}{V} \frac{\int d\mathbf{r}_1 d\mathbf{r}_1' d\mathbf{r}_2 \ldots d\mathbf{r}_N \Psi_G(\mathbf{r}_1, \mathbf{r}_2 \ldots \mathbf{r}_N) \Psi_G(\mathbf{r}_1', \mathbf{r}_2 \ldots \mathbf{r}_N)}{\int d\mathbf{r}_1 d\mathbf{r}_2 \ldots d\mathbf{r}_N \Psi_G^2(\mathbf{r}_1, \mathbf{r}_2 \ldots \mathbf{r}_N)}. \tag{5}$$

For the benefit of the reader, I'll just present here the necessary ingredients needed for a simple proof of the above relation. First the normalized ground state has to be written in second quantized form as

$$|G\rangle = (N! \int d\mathbf{r}_1 \cdots d\mathbf{r}_N \Psi_G^2(\mathbf{r}_1, \cdots \mathbf{r}_N))^{-1/2} \int d\mathbf{r}_1 \cdots d\mathbf{r}_N \Psi_G(\mathbf{r}_1, \cdots, \mathbf{r}_N) \psi^+(\mathbf{r}_1) \cdots \psi^+(\mathbf{r}_N) |0\rangle, \tag{6}$$

where the $\psi^+$'s are local Bose creation operators and $|0\rangle$ is the vacuum (no particle) state. Then the creation operator in the single particle zero momentum state is expressed as

$$a_0^+ = V^{-1/2} \int d\mathbf{r}\, \psi^+(\mathbf{r}) \tag{7}$$

and the condensed fraction is written as

$$\xi = \langle G | a_0^+ a_0 | G \rangle. \tag{8}$$

The rest of the job is done by using the Bose commutation relations

$$[\psi(\mathbf{x}), \psi(\mathbf{y})] = [\psi^+(\mathbf{x}), \psi(\mathbf{y})] = 0 \text{ and } [\psi(\mathbf{x}), \psi^+(\mathbf{y})] = \delta(\mathbf{x} - \mathbf{y}). \tag{9}$$

Consider a classical system, described by a potential energy given by a sum of pair potentials, which at temperature $T$ is in the liquid phase. The corresponding condensed fraction is obviously a functional of the pair potential $\phi$ and a function of the temperature and the density. In the following it will be shown that in the liquid state the condensed fraction is non-vanishing. For the case under consideration the condensed fraction is given by

$$\xi = \int d\mathbf{r}_1 d\mathbf{r}_1' d\mathbf{r}_2 \ldots d\mathbf{r}_N \exp\{-(1/2)\beta\{\sum_{i=2}^{N}[\phi(\mathbf{r}_i - \mathbf{r}_1) + \phi(\mathbf{r}_i - \mathbf{r}_1')] + \sum_{i,j=2}^{N} \phi(\mathbf{r}_i - \mathbf{r}_j)\}\}/Q_N V \ , (10)$$

where $Q_N$ is the $N$ particle partition function. Now, multiply and divide the right hand side of the above by $Q_{N+1}$ and note next that $Q_{N+1}$ can be obtained from an integral very similar to that on the right hand side of equation (10). The difference being that the integrand in $Q_{N+1}$ has an additional factor of $\exp\{-(1/2)\beta\sum_{i=2}^{N}[\phi(\mathbf{r}_i - \mathbf{r}_1) + \phi(\mathbf{r}_i - \mathbf{r}_1') + 2\phi(\mathbf{r}_1 - \mathbf{r}_1')]\}$. Because of the short range of $\phi$ the last term in the exponent can be dropped when integrating over $\mathbf{r}_1$ and $\mathbf{r}_1'$ so, that the final result is

$$\xi = \frac{Q_{N+1}}{VQ_N} \frac{G[(1/2)\beta]}{G[\beta]}, \tag{11}$$

where

$$G(\alpha) = \int d\mathbf{r} \left\langle \exp\{-\alpha \sum_{i=1}^{N-1}[\phi(\mathbf{r}_i) + \phi(\mathbf{r}_i + \mathbf{r})]\} \right\rangle \tag{12}$$

and where the average is taken with respect to the Gibbs distribution at temperature $T$ of a system of $N-1$ particles interacting via the two body potential $\phi$. It is clear that the ratio $Q_{N+1}/VQ_N$ is of order one. The ratio of the $G$'s is also of order one. To see that it will be more convenient to express $G(\alpha)$ in the form

$$G(\alpha) = \int d\mathbf{r} \left\langle \exp\{-\alpha \int d\mathbf{r}' \phi(\mathbf{r}')[\rho(\mathbf{r}') + \rho(\mathbf{r}' - \mathbf{r})]\} \right\rangle, \tag{13}$$

where $\rho(\mathbf{r}) = \sum_{i=1}^{N-1} \delta(\mathbf{r} - \mathbf{r}_i)$ is the particle density. Since the two particle potential is short ranged the expression for G can be simplified to

$$G(\alpha) = V \left\langle \exp\{-2\alpha \int d\mathbf{r}' \phi(\mathbf{r}')\rho(\mathbf{r}')\} \right\rangle . \tag{14}$$

Each of the $G'$ is of order $V$ and so their ratio is of order one. In fact, for the hard sphere system the ratio of the $G$'s is just 1. Thus the condensed fraction is non-vanishing in a classical liquid.

The frozen state, be it a periodic crystal or a glass is characterized by the fact that each particle is localized in the vicinity of some fixed point in space. The analogy with the quantum mechanical Bose system, this time with a system, in which the particles are localized, is still valid, because the localization can be taken into account by a proper modification of the classical distribution function. The problem of the existence of Bose Einstein condensation in a Bose solid was discussed extensively in the past [5-9] and these ideas can be borrowed for the discussion of the frozen state in the classical case through the equivalence between the quantum mechanical ground state $\Psi_G$ and the square root of the classical distribution function. The relevant result for the present discussion is that Bose Einstein condensation does not exist in a Bose system in which the particles are localized. Therefore it does not exist also in the frozen classical system. For the sake of completeness an outline of the argument is given in the following.

A reasonable configuration space distribution function corresponding to the frozen system, $P_f$ is given by

$$P_f = P_S \sum_p L[\mathbf{r}_1, \ldots \mathbf{r}_N; p(\mathbf{R}_1, \ldots \mathbf{R}_N)] , \tag{14}$$

where the sum on $p$ is the sum over permutations of the lattice (either periodic or disordered) sites. The function $L$ is given by

$$L[\mathbf{r}_1, \ldots \mathbf{r}_N; \mathbf{R}_1, \ldots \mathbf{R}_N] = \prod_{i=1}^{N} g(\mathbf{r}_i - \mathbf{R}_i, \mathbf{R}_i), \tag{15}$$

where the function $g(\mathbf{r}, \mathbf{R}_i)$ viewed as a function of $\mathbf{r}$ is one within some compact singly connected region around the origin and vanishes outside it. Its dependence on $\mathbf{R}_i$ denotes that the shape and orientation of the region in which the function differs from zero, depends on the lattice site. (Even for the more familiar ordered lattice such a form is necessary as $P_S$ alone is invariant under translations.) Consider the off diagonal correlation function

$$\Phi(\mathbf{r}) = \frac{\int d\mathbf{r}_1 \ldots d\mathbf{r}_N \Psi_f(\mathbf{r}_1, \ldots \mathbf{r}_N) \Psi_f(\mathbf{r}_1 + \mathbf{r}, \mathbf{r}_2, \ldots \mathbf{r}_N)}{\int d\mathbf{r}_1 \ldots d\mathbf{r}_N \Psi_f^2(\mathbf{r}_1, \ldots \mathbf{r}_N)} , \tag{16}$$

where $\Psi_f = P_f^{1/2}$. The system will exhibit a nonzero condensed fraction if and only if the off diagonal correlation has an infinite range. In order to understand what is going on it is instructive to consider first the case of no overlap. This is the case where $g(\mathbf{r} - \mathbf{R}_i, \mathbf{R}_i) g(\mathbf{r} - \mathbf{R}_j, \mathbf{R}_j) = 0$ for $i \neq j$. Each of the $\Psi$'s in the integrand in the numerator on the right hand side of eq. (14) involves a sum over permutations. The integrand in the numerator on the right hand side of eq. (16) will thus involve a double sum over permutations of terms of the form $L[\mathbf{r}_1, \ldots \mathbf{r}_N; p(\mathbf{R}_1, \ldots \mathbf{R}_N)] L[\mathbf{r}_1 + \mathbf{r}, \ldots \mathbf{r}_N; p'(\mathbf{R}_1, \ldots \mathbf{R}_N)]$. Because of the condition of no overlap applied to $\mathbf{r}_2, \cdots, \mathbf{r}_N$, it is evident that any product of two $L$'s corresponding to two different permutations is identically zero. In case, $p' = p$, it is obvious that for $|\mathbf{r}|$ large enough (larger than the linear size over which $g(\mathbf{r}, p(\mathbf{R}_1))$ is different from zero) the product of $L$'s is again identically zero. Therefore, the off diagonal correlation (16) vanishes identically for finite large enough distances. For small overlap it will never vanish identically but will decay over a finite range, so that the condensed fraction is zero [9]. In reality the overlap is small, describing the fact that a particle cannot escape from the cage consisting of its neighbors. So that condensation does not exist.

The fact that condensation exists in the liquid state and does not exist in the frozen state, implies that a dimensionless parameter, $g$, that measures the glassiness of the liquid can be defined

$$g = 1 - \xi. \tag{17}$$

Thus, the glassiness changes continuously from zero in the extremely dilute gas to one in the frozen state.

Clearly, the attractive part of the interaction is important in the freezing process. Yet what happens when particles interacting via a purely repulsive interaction are squeezed together is also of interest. In the following I'll give two examples in which known results from classical liquid theory will be used to obtain the glassiness in liquids with repulsive interactions.

The first example is that of the one dimensional hard rode problem. It is interesting not only because it can be solved exactly but also, because of its special peculiarities, which result from the one dimensional character of the problem. The one dimensional hard rod problem is soluble and $Q_N$ can be obtained exactly [11, 12],

$$Q_N = [L - (N-1)a]^N, \tag{18}$$

where $L$ is the total length of the system (not periodic) and $a$ the length of a single rod. Using eq. (11) and the fact that the ratio of the $G$'s is one in the case of hard cores, it is easy to obtain the condensed fraction,

$$\xi = (1-x)^{-x/(1-x)}, \tag{19}$$

where $x = Na/L$. This result of a finite condensed fraction for the hard rod case is actually misleading. The reason is that because of the strict hard rod condition, symmetry under permutation is obviously irrelevant because the particles cannot be interchanged dynamically and the initial order of the different particles is preserved. In $Q_N$ all possible orders are taken into account. Thus the dynamically relevant quantity to replace $Q_N$ is $Q_N / N!$. This leads, as expected, to a vanishing condensed fraction. This is a peculiar result for two reasons. First it is specific to the hard core one dimensional system and second the condensed fraction vanishes while the particles are not localized and do not form a glass. The last property is a peculiarity of one dimension. Only in one dimension with a hard core interaction symmetry under permutations can be broken in spite of the fact that the system is not frozen. In higher dimensions, the dynamical irrelevance of symmetry under permutations can follow only from localization. To get some idea of what happens in the liquid phase in three dimensions consider next the hard sphere system. I will use here the Percus Yevick approximation[13], in spite of its known short comings at high densities, because it offers an analytic [14] and reasonable result over a wide range of densities. The PY equation of state is given by

$$p = nkT[1+\eta+\eta^2]/[1-\eta]^3 , \qquad (20)$$

where $\eta$ is expressed in terms of the hard sphere radius $R$ and the particle density $\rho$, $\eta = \pi R^3 \rho/6$. Once the pressure is given $Q_N$ can be obtained readily resulting in a straight forward manner in an expression for the condensed fraction

$$\xi = (1-\eta)^{f(\eta)} \text{ with } f(\eta) = (3/2)[1-1/(1-\eta)^2]. \qquad (21)$$

and the glassiness will be just

$$g = 1-(1-\eta)^{f(y)}. \qquad (22)$$

Clearly, the above expression can not be taken seriously, for relatively high densities but can be expected to be a semi quantitative description up to medium density range ($\eta$ about 0.3-0.4).

Obviously, approximation schemes that improve on PY can be used as well as simulations. It will be of great interest to use established analytical and numerical methods to treat cases with soft interactions including attractive parts. The purpose of the present article is just to introduce the basic idea and hopefully it will trigger such activity in the future. I believe that of particular interest will be the study of the glassiness in liquids near their transition point and in the meta-stable domain of super cooled liquids.

**Acknowledgement:** I would like to thank S. F. Edwards for reading a previous version of this article and for his helpful remarks.